\PassOptionsToPackage{unicode}{hyperref}
\PassOptionsToPackage{hyphens}{url}
\PassOptionsToPackage{dvipsnames,svgnames,x11names}{xcolor}
\documentclass[
]{article}

\usepackage{amsmath,amssymb}
\usepackage{iftex}
\ifPDFTeX
  \usepackage[T1]{fontenc}
  \usepackage[utf8]{inputenc}
  \usepackage{textcomp} 
\else 
  \usepackage{unicode-math}
  \defaultfontfeatures{Scale=MatchLowercase}
  \defaultfontfeatures[\rmfamily]{Ligatures=TeX,Scale=1}
\fi
\usepackage{lmodern}
\ifPDFTeX\else  
\fi
\IfFileExists{upquote.sty}{\usepackage{upquote}}{}
\IfFileExists{microtype.sty}{
  \usepackage[]{microtype}
  \UseMicrotypeSet[protrusion]{basicmath} 
}{}
\makeatletter
\@ifundefined{KOMAClassName}{
  \IfFileExists{parskip.sty}{%
    \usepackage{parskip}
  }{
    \setlength{\parindent}{0pt}
    \setlength{\parskip}{6pt plus 2pt minus 1pt}}
}{
  \KOMAoptions{parskip=half}}
\makeatother
\usepackage{xcolor}
\usepackage[top=30mm,left=30mm,heightrounded]{geometry}
\usepackage{orcidlink}
\usepackage{authblk}
\usepackage{appendix}
\makeatletter
\ifx\paragraph\undefined\else
  \let\oldparagraph\paragraph
  \renewcommand{\paragraph}{
    \@ifstar
      \xxxParagraphStar
      \xxxParagraphNoStar
  }
  \newcommand{\xxxParagraphStar}[1]{\oldparagraph*{#1}\mbox{}}
  \newcommand{\xxxParagraphNoStar}[1]{\oldparagraph{#1}\mbox{}}
\fi
\ifx\subparagraph\undefined\else
  \let\oldsubparagraph\subparagraph
  \renewcommand{\subparagraph}{
    \@ifstar
      \xxxSubParagraphStar
      \xxxSubParagraphNoStar
  }
  \newcommand{\xxxSubParagraphStar}[1]{\oldsubparagraph*{#1}\mbox{}}
  \newcommand{\xxxSubParagraphNoStar}[1]{\oldsubparagraph{#1}\mbox{}}
\fi
\makeatother

\providecommand{\tightlist}{%
  \setlength{\itemsep}{0pt}\setlength{\parskip}{0pt}}\usepackage{longtable,booktabs,array}
\usepackage{calc} 
\usepackage{etoolbox}
\makeatletter
\patchcmd\longtable{\par}{\if@noskipsec\mbox{}\fi\par}{}{}
\makeatother
\IfFileExists{footnotehyper.sty}{\usepackage{footnotehyper}}{\usepackage{footnote}}
\makesavenoteenv{longtable}
\usepackage{graphicx}
\makeatletter
\newsavebox\pandoc@box
\newcommand*\pandocbounded[1]{
  \sbox\pandoc@box{#1}%
  \Gscale@div\@tempa{\textheight}{\dimexpr\ht\pandoc@box+\dp\pandoc@box\relax}%
  \Gscale@div\@tempb{\linewidth}{\wd\pandoc@box}%
  \ifdim\@tempb\p@<\@tempa\p@\let\@tempa\@tempb\fi
  \ifdim\@tempa\p@<\p@\scalebox{\@tempa}{\usebox\pandoc@box}%
  \else\usebox{\pandoc@box}%
  \fi%
}
\def\fps@figure{htbp}
\makeatother
\NewDocumentCommand\citeproctext{}{}

\makeatletter
 \let\@cite@ofmt\@firstofone
 \def\@biblabel#1{}
 \def\@cite#1#2{{#1\if@tempswa , #2\fi}}
\makeatother
\newlength{\cslhangindent}
\newlength{\csllabelwidth}
\newenvironment{CSLReferences}[2] 
 {\begin{list}{}{%
  \setlength{\itemindent}{0pt}
  \setlength{\leftmargin}{0pt}
  \setlength{\parsep}{0pt}
  \ifodd #1
   \setlength{\leftmargin}{\cslhangindent}
   \setlength{\itemindent}{-1\cslhangindent}
  \fi
  \setlength{\itemsep}{#2\baselineskip}}}
 {\end{list}}
\usepackage{calc}

\usepackage{booktabs}
\usepackage{longtable}
\usepackage{array}
\usepackage{multirow}
\usepackage{wrapfig}
\usepackage{float}
\usepackage{colortbl}
\usepackage{pdflscape}
\usepackage{tabu}
\usepackage{threeparttable}
\usepackage{threeparttablex}
\usepackage[normalem]{ulem}
\usepackage{makecell}
\usepackage{xcolor}
\usepackage{graphicx, pdflscape, adjustbox, booktabs, tabularx, array, makecell, pdflscape,longtable, multirow}
\makeatletter
\@ifpackageloaded{caption}{}{\usepackage{caption}}
\AtBeginDocument{%
\ifdefined\contentsname
  \renewcommand*\contentsname{Table of contents}
\else
  \newcommand\contentsname{Table of contents}
\fi
\ifdefined\listfigurename
  \renewcommand*\listfigurename{List of Figures}
\else
  \newcommand\listfigurename{List of Figures}
\fi
\ifdefined\listtablename
  \renewcommand*\listtablename{List of Tables}
\else
  \newcommand\listtablename{List of Tables}
\fi
\ifdefined\figurename
  \renewcommand*\figurename{Figure}
\else
  \newcommand\figurename{Figure}
\fi
\ifdefined\tablename
  \renewcommand*\tablename{Table}
\else
  \newcommand\tablename{Table}
\fi
}
\@ifpackageloaded{float}{}{\usepackage{float}}
\floatstyle{ruled}
\@ifundefined{c@chapter}{\newfloat{codelisting}{h}{lop}}{\newfloat{codelisting}{h}{lop}[chapter]}
\floatname{codelisting}{Listing}

\makeatother
\makeatletter
\@ifpackageloaded{caption}{}{\usepackage{caption}}
\@ifpackageloaded{subcaption}{}{\usepackage{subcaption}}
\makeatother

\usepackage{bookmark}

\IfFileExists{xurl.sty}{\usepackage{xurl}}{} 
\hypersetup{
  pdftitle={Artificial Intelligence in Deliberation},
  colorlinks=true,
  linkcolor={blue},
  filecolor={Maroon},
  citecolor={Blue},
  urlcolor={Blue},
  pdfcreator={LaTeX via pandoc}}

\title{Artificial Intelligence in Deliberation: The AI Penalty and the
Emergence of a New Deliberative Divide}
\author[1]{Andreas Jungherr \orcidlink{0000-0003-2598-2453}}
\author[2]{Adrian Rauchfleisch \orcidlink{0000-0003-1232-083X}}
\affil[1]{University of Bamberg}
\affil[2]{National Taiwan University}
\date{\today}

\begin{document}
\maketitle
\begin{abstract}
Digital deliberation has expanded democratic participation, yet
challenges remain. This includes processing information at scale,
moderating discussions, fact-checking, or attracting people to
participate. Recent advances in artificial intelligence (AI) offer
potential solutions, but public perceptions of AI's role in deliberation
remain underexplored. Beyond efficiency, democratic deliberation is
about voice and recognition. If AI is integrated into deliberation,
public trust, acceptance, and willingness to participate may be
affected. We conducted a preregistered survey experiment with a
representative sample in Germany (n=1850) to examine how information
about AI-enabled deliberation influences willingness to participate and
perceptions of deliberative quality. Respondents were randomly assigned
to treatments that provided them information about deliberative tasks
facilitated by either AI or humans. This allows us to identify causal
effects. Our findings reveal a significant AI-penalty. Participants were
less willing to engage in AI-facilitated deliberation and rated its
quality lower than human-led formats. These effects were moderated by
individual predispositions. Perceptions of AI's societal benefits and
anthropomorphization of AI showed positive interaction effects on
people's interest to participate in AI-enabled deliberative formats and
positive quality assessments, while AI risk assessments showed negative
interactions with information about AI-enabled deliberation. These
results suggest AI-enabled deliberation faces substantial public
skepticism, potentially even introducing a new deliberative divide.
Unlike traditional participation gaps based on education or
demographics, this divide is shaped by attitudes toward AI. As
democratic engagement increasingly moves online, ensuring AI's role in
deliberation does not discourage participation or deepen inequalities
will be a key challenge for future research and policy.
\end{abstract}


\section{Highlights}\label{highlights}

\begin{itemize}
\tightlist
\item
  AI-enabled deliberation reduces willingness to participate in
  deliberative formats.
\item
  Participants rate AI-facilitated deliberation lower in quality than
  human-led formats.
\item
  Political interest and need for cognition shape willingness to
  participate in deliberation.
\item
  Positive attitudes toward AI increase willingness to participate in
  deliberation and quality assessments, while AI risk perceptions lower
  them.
\item
  AI in deliberation introduces a new divide beyond traditional
  political participation gaps driven by AI-related preconceptions.
\end{itemize}

\section{Introduction}\label{introduction}

Digital deliberation has provided a valuable extension of the
deliberative toolbox (Landemore, 2021). Digitally mediated deliberation
has proven helpful to formats like citizen assemblies, consultations,
virtual town halls, or mini-publics. But challenges remain. This
includes surfacing information in large deliberative assemblies,
moderating contributions, fact-checking, or providing summaries. Recent
advances in artificial intelligence (AI) promise to provide solutions to
some of these challenges (Landemore, 2024; Tsai et al., 2024). But for
these promises to manifest, we need to understand how people think about
including AI within deliberative processes.

Democratic deliberation is not only about finding efficient solutions.
Fundamentally, democratic deliberation is also a process providing
participants with voice and recognition in the democratic process
(Bächtiger \& Dryzek, 2024; J. S. Fishkin, 2018; Lafont, 2020; Neblo et
al., 2018). Even if AI might contribute to better and more efficient
ways to run digital deliberation, there remains the risk that the
technical nature of AI, its workings, and associated hopes and fears
come to shape people's willingness to participate in deliberative
formats and their sense of deliberative quality. Understanding AI's
(potential) contributions to deliberation means not only focusing on its
technical and functional contribution to deliberative processes and
outcomes. It is also about understanding its impact on people's
willingness to participate in deliberative formats and opinions on their
quality.

We present a first representative account of the impact that information
about AI-enabled deliberation has on people's willingness to participate
in deliberative formats and their quality assessment of AI-enabled
deliberative formats. We ran a preregistered survey experiment with a
representative sample of respondents in Germany (n=1850). We exposed
respondents to descriptions of a set of tasks that are crucial in the
organization and execution of deliberative formats. In the descriptions,
we randomly varied the assignment of these tasks to AI or human
facilitators. This allows us to identify the causal effect that
information about AI-enabled deliberation has on people's willingness to
participate in deliberation and their assessment of deliberative
quality.

We find a significant AI-penalty on people's willingness to participate
in AI-enabled deliberative formats compared to their willingness to
participate in human-run deliberation. We also see that people assess
the deliberative quality of AI-enabled deliberative formats lower than
formats where humans performed facilitation and moderation tasks. We
find willingness to participate in deliberation to be shaped by
psychological predispositions for politics and deliberation, namely
political interest and the need for cognition. We also see assessments
of AI benefits for society and the tendency to anthropomorphize AI to
interact positively with information about AI-enabled deliberation on
people's willingness to participate in AI-enabled deliberation and with
their quality assessments of AI-enabled deliberative formats.
Conversely, we see that AI risk assessments did interact negatively.

Our findings show that AI-enabled deliberation faces a challenge in
public acceptance. Currently, there is a significant AI-penalty for
AI-enabled deliberation, reducing people's willingness to participate in
deliberative formats and their quality assessment of deliberation.
Additionally, using AI in deliberative formats introduces a new
deliberative divide. Traditional deliberative formats already face the
challenge of motivating and recruiting people with low political
interest and low cognitive propensity for deliberation (Coleman \&
Blumler, 2009; Landemore, 2020). Our findings show that AI-enabled
deliberation introduces another deliberative divide -- one driven by
people's attitudes toward AI. This divide does not run parallel to
well-known divides in political participation, such as educational
attainment or socio-demographic inequalities. But this does not make it
less problematic. If political participation is increasingly digitally
mediated and AI-enabled, people's propensity to engage with digital
technology and trust in AI become factors that shape their participatory
opportunities.

\section{Theory: Modeling willingness to participate in AI-enabled
deliberation and assessments of deliberative
quality}\label{theory-modeling-willingness-to-participate-in-ai-enabled-deliberation-and-assessments-of-deliberative-quality}

Recent advances in generative artificial intelligence (AI) promise
improvements to the functionality and quality of digitally mediated
deliberative processes (Landemore, 2024; Small et al., 2023; Tessler et
al., 2024; Tsai et al., 2024). Advanced capabilities in text analysis,
summation, and generation have raised expectations that AI-enabled
deliberative platforms can support deliberative processes in summary of
information and arguments (Arana-Catania et al., 2021; Bakker et al.,
2022; Chowanda et al., 2017; Small et al., 2023; Tessler et al., 2024),
support deliberative exchanges (Arana-Catania et al., 2021; Argyle et
al., 2023; Dooling \& Febrizio, 2023; J. Fishkin et al., 2019) by
enforcing civility and factuality (Agarwal et al., 2024; J. Fishkin et
al., 2019; Giarelis et al., 2024) and enabling comprehension between
diverse sets of participants (Feng et al., 2023; McKinney, 2024; Small
et al., 2023), and support consensus building and decision making by
mapping and aggregating opinions and preferences of participants
(Arana-Catania et al., 2021; Fish et al., 2024; J. Fishkin et al., 2019;
Gudiño-Rosero et al., 2024; Konya et al., 2023; Small et al., 2023).
Using AI for these tasks promises to offset the persisting challenges of
participant management and information aggregation in large-scale online
deliberation (Landemore, 2024). But while the technical potential for AI
to improve deliberation has been convincingly argued and demonstrated in
prototypes, we do not know how AI-enabled deliberation influences
people's readiness to participate in deliberation and their assessment
of the deliberative quality of AI-enabled processes.

Much of the current discussion of the uses of AI in deliberation focuses
on technical and functional aspects of AI-enabled deliberation. These
arguments foreground information processing challenges in large-scale
digital deliberation contexts. Clearly, AI can provide some solutions
for information processing, discovery, and analysis. Still, technical
utility does not automatically lead to public acceptance. Democratic
deliberation is not only about finding a solution to collective
problems; it is also about providing participants with voice and
recognition in the democratic process (Bächtiger \& Dryzek, 2024; J. S.
Fishkin, 2018; Lafont, 2020; Neblo et al., 2018). While AI might
efficiently support collective decision-making through processing and
aggregating information, its role as mediator or moderator of people's
contributions during the deliberative process can negatively impact
participants' sense of voice and recognition (Alnemr, 2020; Lazar \&
Manuali, 2024). Having a machine perform the tasks of human mediators
and moderators within deliberative processes can thereby negatively
impact peoples' readiness to participate and their sense of deliberative
quality. This underlying tension between the promise of increased
capabilities and fears of technology-driven alienation in AI-enabled
deliberation motivates our first research question.

RQ1: Do people react differently to information about the uses of human
vs AI moderators in deliberative formats?\footnote{This translates into
  two subquestions: a) Does information about the uses of human vs AI
  moderators impact people's willingness to participate in said
  deliberative format? b) Does information about the uses of human vs AI
  moderators impact people's assessment of the deliberative quality of
  said deliberative format? For research question and all hypotheses,
  see preregistration
  \url{https://osf.io/aq42e/?view_only=9f19e324effb42d9a5891ceedaf728fa}.}

The readiness to participate in deliberation has been explained in the
past by a combination of material and cognitive factors (Neblo et al.,
2010). Neblo et al. (2010) built their study on the influential
\emph{civic voluntarism model} that sees conventional political
participation as a function of available resources and psychological
engagement with politics (Schlozman et al., 2018). Following the model,
we account for different types of resources influencing people's
readiness to participate in deliberation. Money and time available for
politics should most strongly influence people's willingness to
participate. Income is comparatively easy to measure, but time is more
difficult to capture. We account for people's full-time employment
status and the number of young children in a household. Both should
negatively impact people's willingness to participate in deliberation by
cutting down on the resource time. Another set of resources accounts for
the necessary civic skills enabling people to participate; this includes
educational attainment as a proxy for early exposure to civic skills and
its necessary building blocks and prior experience with political
participation. To account for the specifics of political deliberation,
we measure people's prior experience with political deliberation. To
identify psychological involvement with politics, we consider
respondents' political interest, political efficacy, and their strength
of party identification.

Neblo et al. (2010) argue that the explanation of the willingness to
participate in deliberative formats needs to consider further
explanatory factors that account for ways that political deliberation
differs from other forms of deliberation. This includes psychological
fit with the process of political deliberation and argument and the
aptitude toward specific forms within the implementation of deliberative
formats. Of the potential measures accounting for psychological aptitude
suggested by Neblo et al. (2010), we consider the need for cognition.

This leads us to the following expectations:

Willingness to participate in deliberation for deliberative formats
moderated by humans and AI will be significantly predicted \ldots{}

H1: \ldots{} by variables from the civic voluntarism model. This
includes variables accounting for resources available for participation
(H1a), and psychological engagement with politics (H1b).

H2: \ldots{}
\emph{psychological predisposition} toward deliberation.

Introducing AI to deliberative formats means introducing a controversial
technological feature. AI is a highly controversial technology with
bifurcated expectations (O'Shaughnessy et al., 2023; Zhang, 2024; Zhang
\& Dafoe, 2019) and strong public skepticism, especially regarding its
uses for politics (Jungherr et al., 2024). Finding AI involved in
deliberation might shift people's willingness to participate and their
assessment of deliberative quality according to their attitudes toward
technology or AI.

While AI is broadly and controversially discussed, it is still an
advanced technology with which many people have no direct contact. For
many people, their general attitudes toward the benefits or risks of
technology are likely to serve as moderators for whether they are
interested in participating in AI-enabled deliberation and the quality
they assign to it. More specifically, if people express benefit or risk
perceptions of AI in general (Bao et al., 2022; Zhang \& Dafoe, 2019),
these attitudes likely also matter for their assessments of AI-enabled
deliberation. Furthermore, people's direct experience with AI, either at
work or in their private life, are also indicators of their openness
toward AI and, therefore, are also likely moderators. Finally, we
consider the tendency of people to anthropomorphize machines -- that is,
to attribute human characteristics to them (Epley et al., 2007; Ikari et
al., 2023; Waytz et al., 2010). This tendency has been shown for people
to be more open toward technology use in other contexts and to be less
critical of it. This leads us to expect AI anthropomorphism (Folk et
al., 2025; Ibrahim et al., 2025) to figure as a moderator for attitudes
toward AI-enabled deliberation.

We expect interaction effects of information about AI or human-enabled
deliberative formats. Here, we expect people's general assessments of
technology benefits to positively interact with information of
AI-enabled deliberation on people's willingness to participate in
deliberation (H3a) and their assessment of deliberative quality (H8a).
Conversely, we expect people's assessment of technology risks to carry
negative interaction effects (H3b, H8b). We expect similar interaction
effects for people's general AI benefit (H4, H9) and risk assessments
(H5, H10). We also expect positive interaction effects with people's
professional and personal experience with AI (H6, H11) and their
tendency to anthropomorphize AI (H7, H12). For a complete list of our
preregistered research question and hypotheses, see
Table~\ref{tbl-hypothese}.

\begin{longtable}[]{@{}
  >{\raggedright\arraybackslash}p{(\linewidth - 2\tabcolsep) * \real{0.5333}}
  >{\raggedright\arraybackslash}p{(\linewidth - 2\tabcolsep) * \real{0.4667}}@{}}
\caption{Hypotheses}\label{tbl-hypothese}\tabularnewline
\toprule\noalign{}
\begin{minipage}[b]{\linewidth}\raggedright
Code
\end{minipage} & \begin{minipage}[b]{\linewidth}\raggedright
Research Question and Hypotheses
\end{minipage} \\
\midrule\noalign{}
\endfirsthead
\toprule\noalign{}
\begin{minipage}[b]{\linewidth}\raggedright
Code
\end{minipage} & \begin{minipage}[b]{\linewidth}\raggedright
Research Question and Hypotheses
\end{minipage} \\
\midrule\noalign{}
\endhead
\bottomrule\noalign{}
\endlastfoot
RQ1 & Do people react differently to information about the uses of human
vs AI moderators in deliberative formats? \\
& Does information about the uses of human vs AI moderators impact
people's \ldots{} \\
RQ1a & \ldots{} \emph{willingness to participate} in said deliberative
format? \\
RQ1b & \ldots{} assessment of the \emph{deliberative quality} of said
deliberative format? \\
& \\
& \emph{Willingness to participate} in deliberation for both
deliberative formats moderated by humans and AI will be significantly
predicted \ldots{} \\
H1a & \\
H1b & \ldots{} psychological engagement with politics. \\
H2 & \ldots{} psychological predisposition toward deliberation. \\
& \\
& The effect of information about AI-enabled moderation on the
\emph{readiness to participate} in deliberative formats will positively
interact with \ldots{} \\
H3a & \ldots{} high perceptions of technology's benefits in daily life,
and \\
H3b & \ldots{} low perceptions of technology's risks in daily life. \\
H4 & \ldots{} high perceptions of AI benefits in general. \\
H5 & \ldots{} low perceptions of AI risks in general. \\
H6 & \ldots{} prior AI experience. \\
H7 & \ldots{} high levels of AI-anthropomorphism. \\
& \\
& The effect of information about AI-enabled moderation on positive
\emph{quality assessments of deliberative formats} will positively
interact with \ldots{} \\
H8a & \ldots{} high perceptions of technology's benefits in daily life,
and \\
H8b & \ldots{} low perceptions of technology's risks in daily life. \\
H9 & \ldots{} high perceptions of AI benefits in general. \\
H10 & \ldots{} low perceptions of AI risks in general. \\
H11 & \ldots{} prior AI experience. \\
H12 & \ldots{} high levels of AI-anthropomorphism. \\
\end{longtable}

\section{Data \& Methods}\label{data-methods}

We test the causal effects of learning about different forms of
AI-enabled deliberation. We ran a preregistered survey experiment with
members of an online panel provided by the market and opinion research
company \emph{Ipsos}. We used quotas on age, gender, region, and
education to realize a sample representative of the German population 18
years and older (See Supplementary Materials for details on sampling).
Our final sample consisted of 1850 participants, after removing 97
participants who failed at least two out of three attention
checks.\footnote{See Supplementary Materials for details on exclusions.}.
A prior power analysis indicated a power of 1 for 1800 respondents or
more for the interaction effects in our experiment.\footnote{For R-code
  for the power simulation see preregistration.} Before running the
survey, we registered our research design, analysis plan, and hypotheses
about outcomes.\footnote{For preregistration see
  \url{https://osf.io/aq42e/?view_only=9f19e324effb42d9a5891ceedaf728fa}.
  We only deviate from the preregistration in the measurement of
  education. Other than preregistered, we count BA attainment or higher
  as high education.} The data collection ran between February 11 and
February 18, 2025.

We presented respondents with descriptions of a random selection of a
set of 12 deliberation tasks. This set of tasks captured different
elements that are important to the successful design and execution of
deliberation formats. This includes recruitment of members and topic
selection (i.e., balanced participation, topic identification),
AI-enabled information and argument processing (i.e., content
summarization, argument highlighting, argument structuring, opinion
aggregation), AI-enabled discussion moderation (i.e., fact checking,
disruption detection, tone \& authenticity), and facilitation (i.e.,
gamification, translation support, role-play empathy).\footnote{For a
  detailed list of tasks see Supplementary Materials on Tasks.}

We showed each respondent six randomly drawn deliberation tasks out of a
total of 12. We randomly attributed each task to a human or AI
facilitator. Each respondent saw three tasks attributed to humans and
three tasks attributed to AI. We then measured the effects on two
dependent variables: respondents' willingness to participate in
deliberation and their assessment of deliberative quality.

To identify the effects of AI-enabled deliberation on deliberative
intent, we use a variant of an item previously used by Neblo et al.
(2010).\footnote{For a full list of all measures including variables,
  question wordings, answer options, and key diagnostics see
  Supplementary Materials on Measures.} After showing respondents a
description of an AI- or human-enabled deliberation task within a
deliberative format, we ask: ``If you had the chance to participate in
such a session, how interested do you think you would be in doing
so?''\footnote{Here, we present English translations of the questions we
  ran in our questionnaire in German. For the original German wording of
  our questionnaire see our preregistration:
  \url{https://osf.io/aq42e/?view_only=9f19e324effb42d9a5891ceedaf728fa}.}

To measure the impact of information about AI-enabled deliberation on
people's assessment of deliberative quality, we combine answers to five
questions that capture different aspects of deliberative quality in an
index of deliberative quality. After showing respondents a description
of an AI- or human-enabled task within a deliberative format, we ask
whether respondents agree or disagree with the following statements: (1)
``This makes it difficult for people like me to get their voices heard
{[}inverted{]}''; (2) ``This makes it easy to identify the best policy
solution''; (3) ``This ensures a fair discussion and policy process'';
(4) ``I trust this process''; and (5) ``This will give space for a
plurality of diverse views on the issue''.

Our independent variables are a set of established and new measurements.
We measure the need for cognition with items proposed and validated by
Matthes (2006). We asked respondents whether they agreed or disagreed
with the following statements: (1) ``I find satisfaction in deliberating
hard and for long hours''; (2) ``The notion of thinking abstractly is
appealing to me;'' (3) ``I really enjoy a task that involves coming up
with new solutions to problems;'' and (4) ``I like tasks that require
much thought and mental effort.''

We are additionally interested in how people's general sense of
technology benefits and risks influences their perceptions of AI-enabled
deliberation. We designed two new scales to measure people's sense of
technology's impact on their daily lives.\footnote{For a validation of
  both scales see Supplementary Materials on Validation.} To assess
people's sense of technological benefits, we asked for their agreement
or disagreements with the following three statements: (1) ``I feel more
in control of my daily activities thanks to technology;'' (2)
``Technology helps me manage my time more effectively;'' and (3) I feel
more productive in both my personal and professional life because of
technology. To measure people's sense of technology risks, we asked for
their agreement or disagreements with the following three statements:
(1) ``Technology increases my stress levels rather than making life
easier;'' (2) ``Technology reduces the quality of face-to-face
interactions with others;'' and (3) ``Technology contributes to
distractions and reduces my ability to focus.''

Moving closer to our dependent variables, we ask people for their
general sense of AI benefits and risks. We measure this by two lists of
items that address perceived benefits and risks of AI in different
societal areas: economy, international security, state capacity,
election campaigns, and existential threats. We chose these areas by
drawing on research on AI risk and benefit perceptions (Bao et al.,
2022; Zhang \& Dafoe, 2019) and attitudes toward technology (Binder et
al., 2012; Siegrist \& Visschers, 2013). By querying respondents for
agreement and disagreement with these statements, we develop an index of
their general attitudes toward AI benefits and risks.

To account for respondents' experience with AI, we ask how frequently
they are using AI-supported applications in their (1) ``professional or
work environment'' and (2) ``personal and spare time''. We combine the
two items into a mean index.

Finally, we measure respondents' inclination to project human faculties
onto AI, we build on items proposed by Folk et al. (2025). From their
original measurement scale, we identified four items with high
reliability, adjusted their wording slightly and used them to build our
index of AI anthropomorphism. Our adjusted items are: (1) ``AI has the
potential to develop a sense of humor;'' (2) ``An AI can have a unique
personality;'' (3) ``AI has the capacity to develop consciousness;'' and
(4) ``AI can develop a sense of morals and ethics.''

In Table~\ref{tbl-variables_overview}, we document descriptive
statistics for dependent and independent variables, linked to
preregistered research questions and hypotheses.

\begin{longtable}[]{@{}
  >{\raggedright\arraybackslash}p{(\linewidth - 4\tabcolsep) * \real{0.8000}}
  >{\raggedright\arraybackslash}p{(\linewidth - 4\tabcolsep) * \real{0.1263}}
  >{\raggedright\arraybackslash}p{(\linewidth - 4\tabcolsep) * \real{0.0737}}@{}}
\caption{Descriptive statistics for dependent and indeptendet
variables}\label{tbl-variables_overview}\tabularnewline
\toprule\noalign{}
\begin{minipage}[b]{\linewidth}\raggedright
Variable
\end{minipage} & \begin{minipage}[b]{\linewidth}\raggedright
M (SD)
\end{minipage} & \begin{minipage}[b]{\linewidth}\raggedright
n
\end{minipage} \\
\midrule\noalign{}
\endfirsthead
\toprule\noalign{}
\begin{minipage}[b]{\linewidth}\raggedright
Variable
\end{minipage} & \begin{minipage}[b]{\linewidth}\raggedright
M (SD)
\end{minipage} & \begin{minipage}[b]{\linewidth}\raggedright
n
\end{minipage} \\
\midrule\noalign{}
\endhead
\bottomrule\noalign{}
\endlastfoot
RQ1a/H1-H7: Willingness to participate in deliberation & 3.95 (1.98) &
11100 \\
RQ1b/H8-H12: Deliberative quality (5 items, $\alpha$ = 0.73-0.84) & 4.16 (1.29)
& 11100 \\
H1a: Income & 8.52 (2.85) & 1716 \\
H1a: Employment (full-time=1) & 0.49 & 1823 \\
H1a: Children, 12 or younger & 0.27 (0.82) & 1721 \\
H1a: Education (University degree incl.~Bachelor = 1) & 0.23 & 1850 \\
H1a: Past participatory activity & 0.68 (1.08) & 1641 \\
H1a: Past deliberative activity & 0.39 (0.90) & 1715 \\
H1b: Political interest & 4.91 (1.77) & 1850 \\
H1b: Political efficacy (2 items, Spearman-Brown = 0.67) & 3.34 (1.57) &
1850 \\
H1b: PID strength & 4.55 (1.95) & 1850 \\
H2: Need for cognition (4 items, $\alpha$ = 0.9) & 4.72 (1.38) & 1850 \\
H3a/H8a: Technology benefits (3 items, $\alpha$ = 0.86) & 4.63 (1.38) & 1850 \\
H3b/H8b: Technology risks (3 items, $\alpha$ = 0.76) & 3.65 (1.43) & 1850 \\
H4/H9: AI benefits (5 items, $\alpha$ = 0.87) & 3.76 (1.35) & 1850 \\
H5/H10: AI risks (5 items, $\alpha$ = 0.83) & 4.58 (1.33) & 1850 \\
H6/H11: AI experience (2 items, $\alpha$ = 0.77) & 2.72 (1.66) & 1850 \\
H7/H12: AI anthropomorphism (4 items, $\alpha$ = 0.88) & 3.18 (1.54) & 1850 \\
\end{longtable}

We estimated two multilevel models with varying intercepts for
participants and cases to test our research questions and hypotheses. We
use the varying intercepts to account for the nested data. Each
participant saw six different tasks out of a total of twelve tasks (we
have between 461 and 464 observations for each specific task). Thus, the
total number of observations for the models is 11,000. We estimated two
models, one for each dependent variable, that included all the
independent variables and the preregistered co-variates age and gender
(for the complete models and details on data imputation, see
Supplementary Materials). The independent variables used as interaction
terms (H3-H12) were all mean-centered before including them in the
model. Furthermore, as specified in the preregistration, we did data
imputation for predictors that allowed non-responses (i.e., income or
employment status) by following the procedures recommended in the
literature (van Buuren, 2012/2018; van Buuren \& Groothuis-Oudshoorn,
2011).

\section{Results}\label{results}

\subsection{AI-penalty in
deliberation}\label{ai-penalty-in-deliberation}

\begin{figure}

\centering{

\pandocbounded{\includegraphics[keepaspectratio]{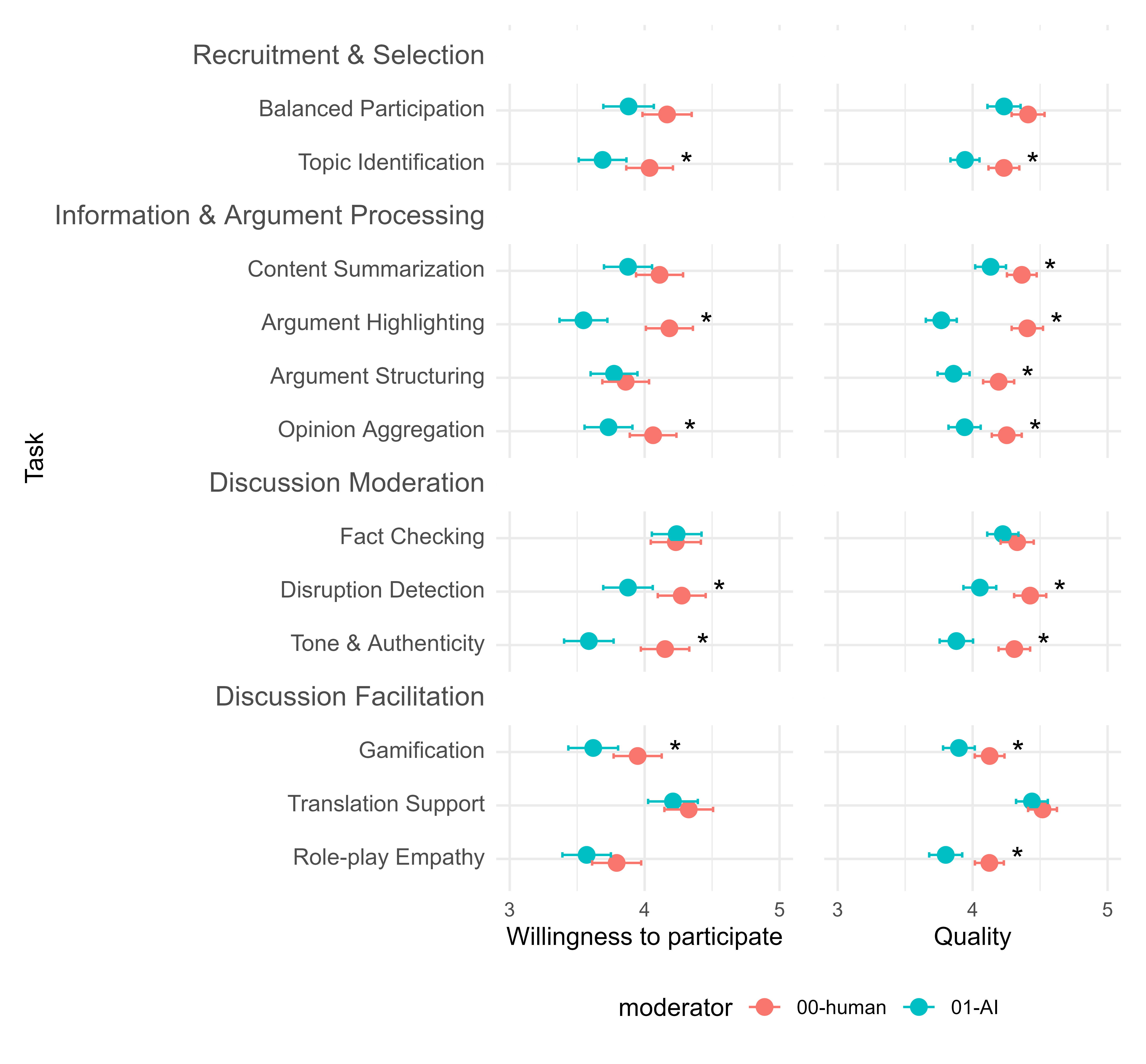}}

}

\caption{\label{fig-tasks}Reaction to deliberative tasks performed by AI
or humans. Points show means with 95\% confidence intervals; stars
indicate significant mean difference after correcting alpha values with
the Benjamini-Hochberg procedure.}

\end{figure}

A core question proponents of AI-enabled features in deliberative
formats face is how people think about AI features in deliberation and
whether they treat AI-enabled deliberative formats differently from
those facilitated by humans (RQ1). In Figure~\ref{fig-tasks} we show the
results of a not preregistered exploratory analysis that informs our
understanding on how people think about AI- and human-enabled
deliberation. We plot respondents' reactions to information about
deliberative tasks performed by AI or humans.\footnote{Values are
  reported after alpha correction (Benjamini \& Hochberg, 1995). Stars
  indicate significant differences identified through t-tests.}

Figure~\ref{fig-tasks} plots people's willingness to participate in
deliberative formats in which specific tasks are performed by AI or by
humans and their assessments of deliberative quality. We see for many
tasks that people prefer deliberative formats in which humans perform
crucial tasks over those enabled by AI. Interestingly, people dislike
many tasks that the literature tends to identify as especially promising
contributions of AI to deliberative formats, namely AI-enabled
information and argument processing (i.e., content summarization,
argument highlighting, argument structuring, opinion
aggregation)(Berliner, 2024; Landemore, 2024; Tessler et al., 2024; Tsai
et al., 2024). Meanwhile, others for which AI faces obvious challenges
-- such as fact-checking --, are seen as unproblematic. This shows that
public expectations for the contribution of AI to deliberation and those
by academics and practitioners deviate.

By comparing respondents' answers across all deliberative tasks
attributed to AI or human moderation, we can identify the causal effect
of information about AI-enabled deliberation compared to information
about human facilitation. We see whether respondents were more or less
willing to participate in AI-enabled deliberation compared to
human-enabled deliberation (RQ1a) and attribute higher or lower
deliberative quality (RQ1b) to either format.

AI-enabled deliberative formats generally meet with significantly lower
willingness to participate by our respondents (b = -0.30, 95\% CI
{[}-0.34, -0.26{]}, p \textless{} .001). Similarly, respondents rated
the deliberative quality of AI-enabled formats lower than those in which
humans were assigned moderation tasks (b = -0.29, 95\% CI {[}-0.32,
-0.26{]}, p = .001). AI-enabled deliberation, therefore, carries an
AI-penalty for both willingness to participate and anticipatory
assessments of deliberative quality.

\subsection{Who is willing to participate in deliberative
formats?}\label{who-is-willing-to-participate-in-deliberative-formats}

\begin{figure}

\centering{

\pandocbounded{\includegraphics[keepaspectratio]{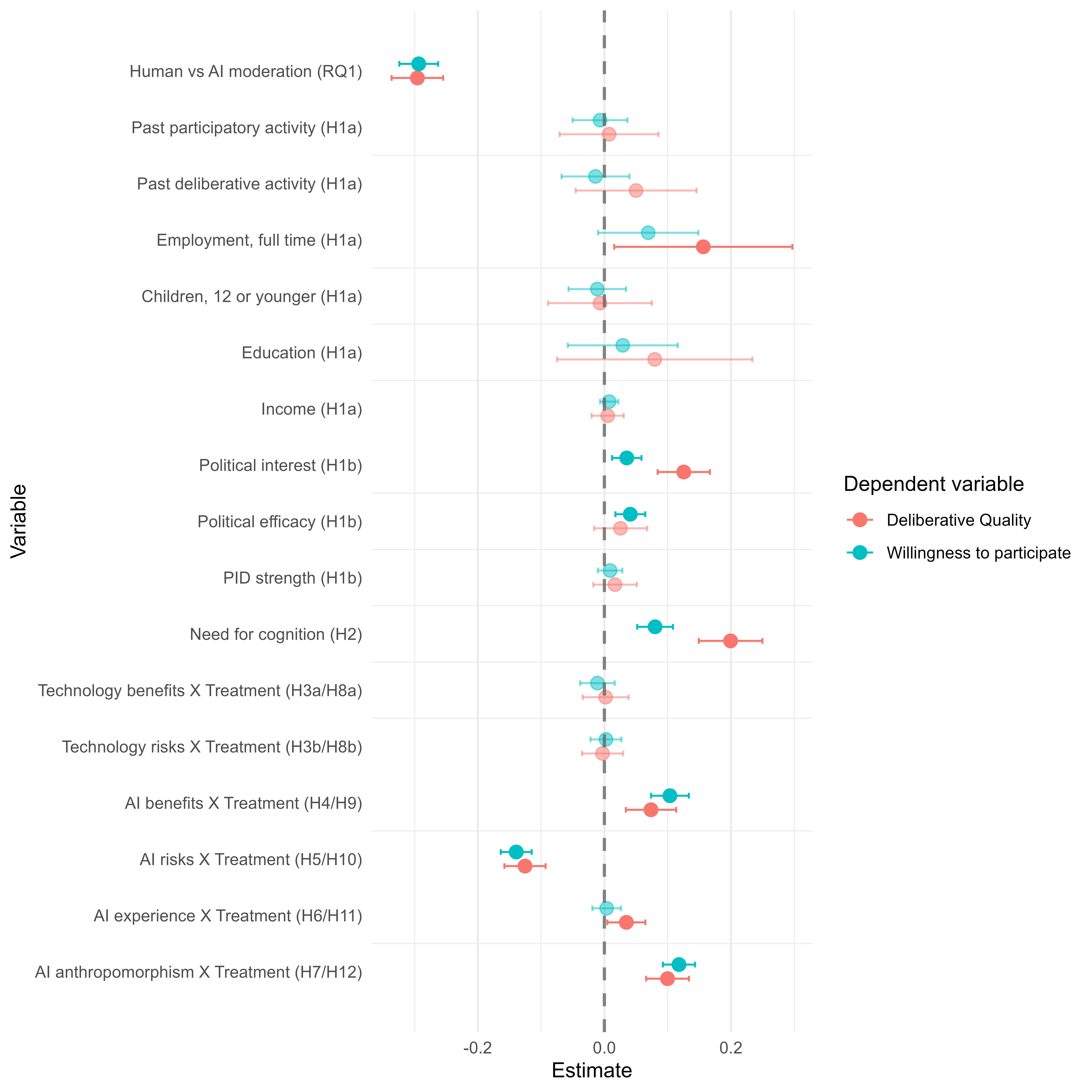}}

}

\caption{\label{fig-tree_plot}Models explaining willingness to
participate in deliberation and quality assessments.}

\end{figure}

A core challenge to the designers and organizers of deliberative formats
is the recruitment of participants, so they constitute a representative
subset of the population. This is made difficult by different population
subsets having different propensities to participate in deliberation,
with those with lower levels of political involvement and fewer
available socio-economic resources tending to be less willing to engage
(Jacobs et al., 2009; Neblo et al., 2010). The impact of AI on people's
heightened or lowered willingness to participate in deliberative formats
can strengthen or weaken the legitimacy of AI-enabled deliberation. We
plot results to our preregistered model\footnote{Figure~\ref{fig-tree_plot}
  documents coefficients to variables linked to theoretically informed
  preregistered hypotheses. Full model specifications are available in
  Supplementary Materials on Model Results.} in
Figure~\ref{fig-tree_plot}.

Regarding the variables from the civic voluntarism model (Schlozman et
al., 2018) our data does not support the expected influence of people's
resources available for political participation (H1a). People's income
(b = 0.01, 95\% CI {[}-0.02, 0.03{]}, p = 0.692), number of children in
the house aged 12 or younger (b = -0.01, 95\% CI {[}-0.09, 0.07{]}, p =
0.868), educational attainment (b = 0.08, 95\% CI {[}-0.07, 0.23{]}, p =
0.312), or even past participatory (b = 0.01, 95\% CI {[}-0.07, 0.09{]},
p = 0.853) or deliberative activity (b = 0.05, 95\% CI {[}-0.05,
0.15{]}, p = 0.304) are all not significant for people's willingness to
participate in deliberative formats enabled by AI or humans.
Furthermore, in contradiction to the underlying model, people's
full-time employment status had a positive effect on their expressed
interest to participate in deliberation (b = 0.16, 95\% CI {[}0.02,
0.30{]}, p = 0.030). In combination, material resources -- be it money,
time, or education -- did not influence people's willingness to
participate in deliberative formats.

The other variable from the civil voluntarism model, psychological
engagement with politics (H1b), had more of the expected impact. As
expected, political interest had a positive effect on people's
willingness to participate in deliberation (b = 0.13, 95\% CI {[}0.08,
0.17{]}, p \textless{} .001). Still, for other indicators of
psychological engagement with politics -- political efficacy (b = 0.03,
95\% CI {[}-0.02, 0.07{]}, p = 0.234) and PID strength (b = 0.02, 95\%
CI {[}-0.02, 0.05{]}, p = 0.340) -- the results were not significant.

Following (Neblo et al., 2010), we also account for people's
psychological predisposition toward deliberation (H2). As expected, we
found that people's expressed need for cognition had a positive effect
on their willingness to participate in deliberation (b = 0.20, 95\% CI
{[}0.15, 0.25{]}, p \textless{} .001).

Our findings indicate that people's willingness to participate in
deliberation largely follows expected patterns. This is true for
cognitive factors, such as political interest and need for cognition.
Material factors may have underperformed because we measured only
expressed willingness to participate rather than actual follow‐through,
and scarce resources like time or money likely matter more when people
must actually spend them -- playing a lesser role in hypothetical
scenarios.

\subsection{Attitudes toward AI mitigate
AI-penalty}\label{attitudes-toward-ai-mitigate-ai-penalty}

In our models explaining people's willingness to participate in
deliberation and their assessment of deliberative quality, we introduce
an additional set of factors accounting for the influence of people's
prior attitudes toward technology and AI (see
Figure~\ref{fig-tree_plot}). Our preregistered model expects that
people's attitudes toward technology and AI shape the effect that
information about AI-enabled deliberation has on people's willingness to
participate in deliberative formats (H3-H7) and on their assessment of
deliberative quality (H8-H12).

Our data do not support H3a/b and H8a/b. We do not find significant
interaction effects of general technology attitudes for both willigness
to participate (technology benefit: b = 0.00, 95\% CI {[}-0.03, 0.04{]},
p = 0.915; technology risk: b = 0.00, 95\% CI {[}-0.04, 0.03{]}, p =
0.861) or assessment of deliberative quality (technology benefit: b =
-0.01, 95\% CI {[}-0.04, 0.02{]}, p = 0.430; technology risk: b = 0.00,
95\% CI {[}-0.02, 0.03{]}, p = 0.843).

\begin{figure}

\centering{

\pandocbounded{\includegraphics[keepaspectratio]{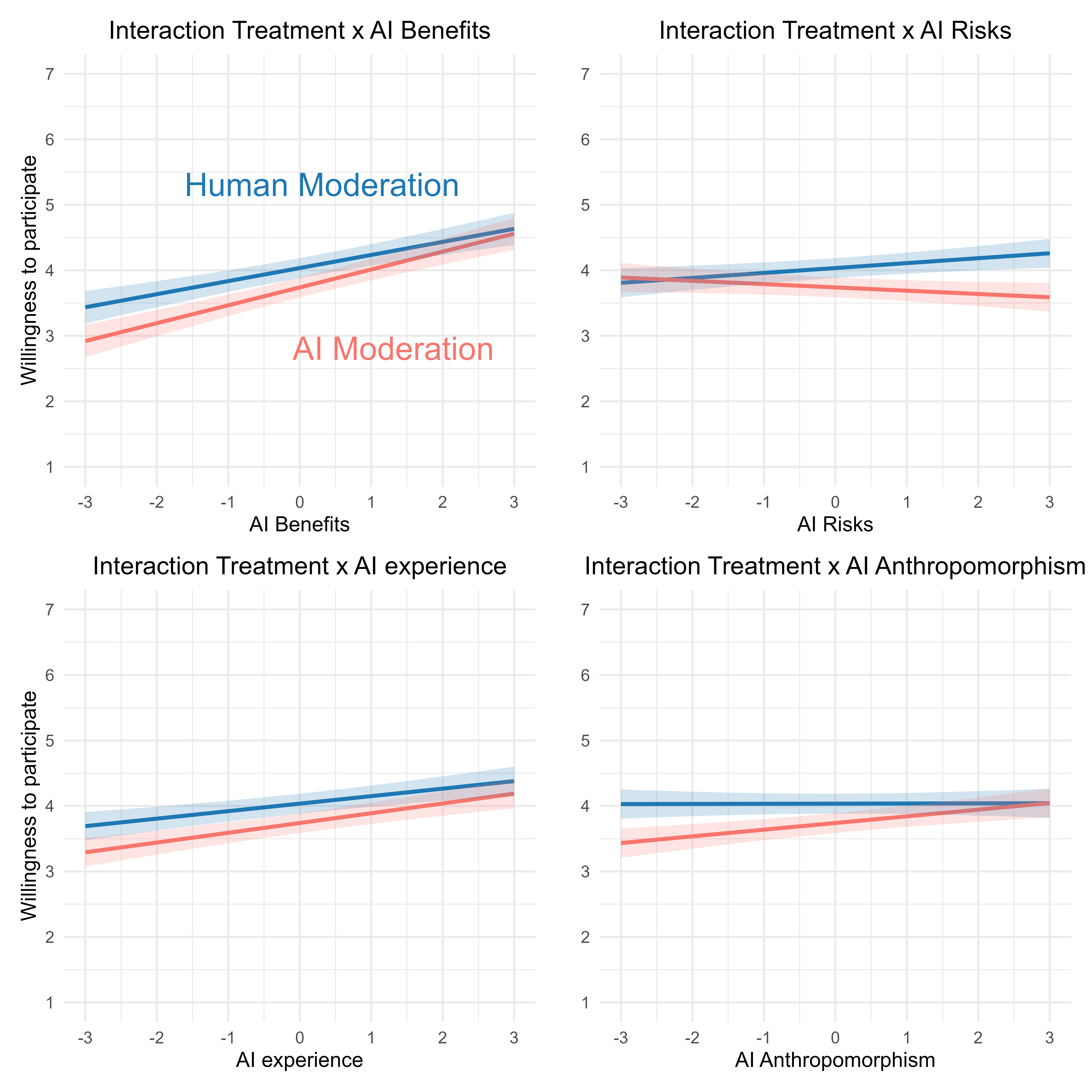}}

}

\caption{\label{fig-interaction_interest}Interaction effects on interest
to participate in deliberation.}

\end{figure}%

In Figure~\ref{fig-interaction_interest}, we plot the interactions
between treatment assignments to information about AI versus
human-enabled deliberative formats with assessments of AI benefits,
risks, prior AI experience, and the tendency to anthropomorphize AI,
with willingness to participate as the dependent variable. Respondents
expressing low assessments of AI benefits express less willingness to
participate in AI-enabled deliberation than in human-enabled
deliberation. Conversely, people who express a high sense of AI benefits
express higher willingness to participate in AI-enabled deliberation
than human-enabled deliberative formats (b = 0.07, 95\% CI {[}0.03,
0.11{]}, p \textless{} .001).

The converse pattern also holds, as shown with attitudes toward AI risk.
Here, we find people with a low sense of AI risks are more willing to
engage in deliberation than those with a heightened sense of AI risk (b
= -0.13, 95\% CI {[}-0.16, -0.09{]}, p \textless{} .001). For prior AI
experience (b = 0.03, 95\% CI {[}0.00, 0.06{]}, p = 0.023) and the
tendency to anthropomorphize AI (b = 0.10, 95\% CI {[}0.07, 0.13{]}, p
\textless{} .001) we observe a positive interaction. Thus, H4-H7 are all
supported by our data.

\begin{figure}

\centering{

\pandocbounded{\includegraphics[keepaspectratio]{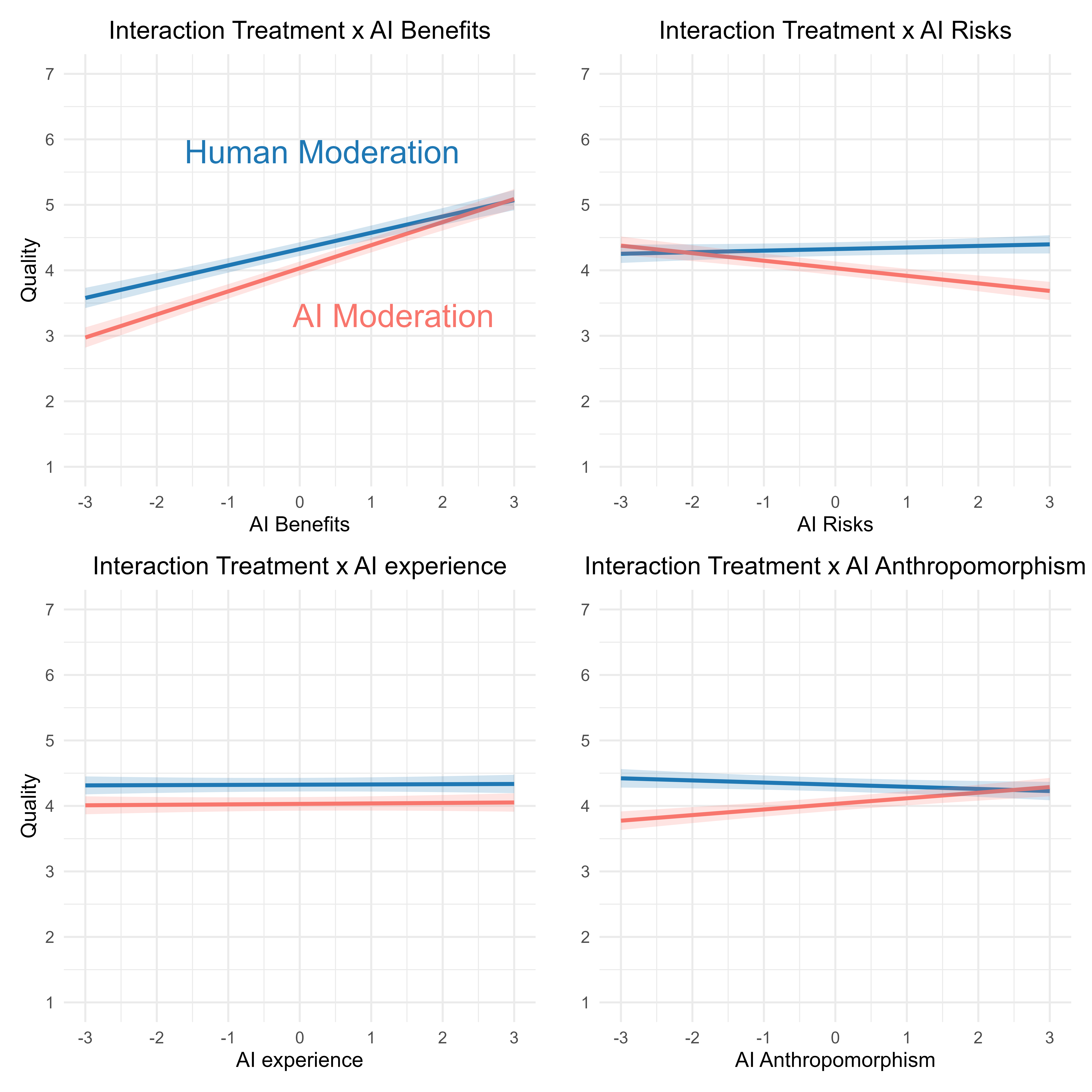}}

}

\caption{\label{fig-interaction_quality}Interaction effects on
assessments of deliberative quality.}

\end{figure}%

Figure~\ref{fig-interaction_quality} shows similar patterns for the
interactions with deliberative quality as the dependent variable.
Respondents expressing low assessments of AI benefits express lower
assessments of deliberative quality for AI-enabled deliberation than for
human-enabled deliberation. Conversely, people who express a high sense
of AI benefits express higher quality assessments for AI-enabled
deliberation than human-enabled deliberative formats (b = 0.10, 95\% CI
{[}0.07, 0.13{]}, p = \textless{} .001). The converse pattern also
holds, as shown with attitudes toward AI risk. Here, we see people with
a low sense of AI risks more likely to express higher deliberative
quality for AI-enabled deliberation than those with a heightened sense
of AI risk (b = -0.14, 95\% CI {[}-0.16, -0.11{]}, p = \textless{}
.001). For H 12, the tendency to anthropomorphize AI (b = 0.12, 95\% CI
{[}0.09, 0.14{]}, p = \textless{} .001), our data supports the
hypotheses as we observe a positive interaction. However, for prior AI
experience, our data does not support H11 (b = 0.00, 95\% CI {[}-0.02,
0.03{]}, p = 0.759).

Generally, the interactions illustrate the mechanism behind the
AI-penalty. People with comparatively high levels of AI benefit
assessments, AI experience (in the case of willingness to participate),
AI anthropomorphism, and comparatively low levels of AI risk
assessments, treat AI-enabled deliberation by and large like
human-enabled deliberation. There is no AI-penalty for these
respondents. All other respondents treat AI-enabled deliberation more
skeptical than human-enabled deliberation. For these respondents, we
find an AI-penalty in deliberation.

\section{Discussion}\label{discussion}

We tested the effect on people being informed about the use of AI for
the performance of important tasks within deliberative formats. Our
findings showed a clear AI-penalty. Respondents expressed less
willingness to participate in AI-enabled deliberation and lower
assessments of deliberative quality, compared to human-enabled
deliberation. This AI-penalty is especially pronounced for people who
are skeptical about AI's role in society. Conversely, people who tended
to see societal benefits in AI, low AI risks, and who anthropomorphized
AI, we found little to no AI-penalty.

These findings are important as they point to currently little discussed
risks to the integration of AI in deliberative formats. If AI reduces
people's willingness to participate in deliberative formats, the
potentially helpful aspects of AI to the conduct of digitally mediated
deliberation are offset. Looking only at potential efficiency gains
within deliberative formats while ignoring the attitudinal and
motivational effects AI uses in deliberation carry, risks missing an
important factor. This is especially true if, as we show, the AI-penalty
is not uniformly distributed through society but instead follows prior
attitudes toward AI. This threatens to introduce a new dimension of
deliberative inequality by having people skeptical about AI to withdraw
from AI-enabled deliberative and participatory opportunities. Proponents
of AI-enabled deliberation, and AI use in democratic practice in
general, need to take this risk seriously and propose mechanisms to
offset this new technology-driven participatory divide.

Our findings also inform discussions about AI-enabled deliberation and
participation more broadly. In explaining people's willingness to
participate in deliberative formats -- AI-enabled or reliant on humans
-- factors based on the civic voluntarism model (Schlozman et al., 2018)
accounting for material resources did not contribute to the explanation
as expected. In contrast, we found cognitive drivers of unequal
participatory interest to hold as expected. As expected, both political
interest (Schlozman et al., 2018) and need for cognition (Neblo et al.,
2010) explained willingness to participate in deliberation. One reason
for the surprising underperformance of material factors could be that we
only queried people for their interest in participation and did not
measure their actual followthrough. Scarceness of material resources,
such as time or money, might come only to figure when people face having
actually to spend time for deliberation while not playing an important
role at earlier state of hypothetical commitment (Neblo et al., 2010).

We also found, somewhat surprisingly, that in their responses to
information about AI-enabled deliberation for tasks associated with
information and argument processing, people appeared especially
hesitant. This stands in contrast with discussion among academics and
practitioners, who tend to see the strongest contribution of AI to
deliberation in information processing and argument mapping (Landemore,
2024; Tessler et al., 2024; Tsai et al., 2024). While people were
skeptical of AI for deliberative information processing and argument
mapping, for fact-checking they rated it similar to human facilitators.
This is interesting, as arguably AI -- at least in its current state --
should be less suited to tasks of fact identification and correction
than information processing or argument mapping. This indicates that
academics and practitioners should not only focus on technically
evaluating AI's information processing and argument mapping tasks but
should also account for people's reactions and assessments. Views of
academics, practitioners, and the public on benefits and risks of AI in
deliberation might deviate more strongly than academics might think.

Even more fundamentally, the identification of an AI-penalty in
deliberation contradicts other recent findings that pointed to people
being open to AI-enabled deliberation as well as expressing higher
satisfaction with AI-enabled facilitation than those by humans (Tessler
et al., 2024). One reason for this discrepancy could be that for our
respondents AI-enabled interventions remain hypothetical, as they do not
experience said interventions directly but are only informed about them.
Indeed, experiencing AI might make people more supportive of it. But our
findings indicate that we should not take AI-acceptance as a given. When
asked for their interest to participate in deliberation, most people
will not yet have direct experience with AI in deliberative formats. The
initial skepticism identified by us has therefore high ecological
validity as the information our respondents have corresponds with that
regular people have when deciding on whether to participate in
AI-enabled deliberation or not. And here, the AI-penalty is real. If we
are interested in increasing participation in deliberative formats, we
need not only focus on people who are willing to participate anyway. The
important challenge lies in including people who are not interested in
participating in the first place. And here our findings indicate that
AI-enabled deliberation might increase hesitancy to participate along
new attitudinal dimensions and therefore weaken the representativeness
and legitimacy of deliberative formats.

Of course, our study comes with limitations. For one, our findings are
based on German respondents. While general psychological processes and
information effects can be expected to be similar across countries with
comparable cultural, political, and technological constellations, it is
unclear if this is the case for AI. Countries might vary regarding their
cultural affinity toward AI. Accordingly, the penalty identified by us
might also vary. Accordingly, future research should focus on
establishing comparative evidence on the AI-penalty.

There are also potential temporal limitations to our study. We show that
willingness to participate in and quality assessments of AI-enabled
deliberation varies based on attitudes toward AI. But how these
attitudes emerge and how they are distributed across societies we do not
know. Nor do we know the prospective development of public opinion on
AI. AI attitudes could turn sour over time, once job replacement or
runaway technological change are more clearly felt across societies.
Alternatively, AI attitudes could turn more positive once people
experience AI-driven increases in welfare and quality of life. Or AI
could normalize. This would make AI-enabled deliberation not special
anymore and, in turn, lead to AI attitudes not playing much of a role in
the readiness to participate in AI-enabled deliberation anymore.
Potential temporal variations should, therefore, also be taken into
account by future research.

Overall, we see that attitudes toward AI matter for both participatory
willingness and assessments of deliberative quality once AI is
introduced in deliberative formats. This is important to keep in mind
for designers of deliberative environments so that they accidentally do
not introduce new dimensions of unequal participation or quality
assessment to a process that already suffers from challenges of unequal
recruitment and associated legitimacy challenges based on unequal
representation. AI might make digitally mediated deliberation more
efficient, but it carries new challenges of public trust and willingness
to engage. This challenge is likely to increase if public skepticism
toward AI rises.

\section{Data Availability}\label{data-availability}

Preregistration information and data are available at the project's OSF
repository:

\begin{itemize}
\tightlist
\item
  Prereg:
  \url{https://osf.io/aq42e/?view_only=9f19e324effb42d9a5891ceedaf728fa}
\item
  Data: Available on OSF
  \url{https://osf.io/5346r/files/osfstorage?view_only=e204fb9a9d6e4af187c7d95934f92cf0}
\end{itemize}

\section{Code Availability}\label{code-availability}

Replication code and data are available at the project's OSF repository:

\begin{itemize}
\tightlist
\item
  Replication code and data: Available on OSF
  \url{https://osf.io/5346r/files/osfstorage?view_only=e204fb9a9d6e4af187c7d95934f92cf0}
\end{itemize}

\section{Acknowledgements}\label{acknowledgements}

A. Jungherr and A. Rauchfleisch contributed equally to the project. The
experiments were preregistered at OSF
\url{https://osf.io/aq42e/?view_only=9f19e324effb42d9a5891ceedaf728fa}
and approved by the IRB at University of Bamberg. Correspondence and
requests for materials should be addressed to A. Jungherr.

\section{Funding sources}\label{funding-sources}

A. Jungherr's contribution is funded by the European Union under grant
agreement No.~101178806. Views and opinions expressed are however those
of the author(s) only and do not necessarily reflect those of the
European Union or European Research Executive Agency. Neither the
European Union nor the granting authority can be held responsible for
them. A. Rauchfleisch's work was supported by the National Science and
Technology Council, Taiwan (R.O.C) (Grant No 113-2628-H-002-018-).

\section{Declaration of interests}\label{declaration-of-interests}

A. Jungherr and A. Rauchfleisch declare no potential competing interests
with respect to the research, authorship, and/or publication of this
article.

\section{Author Information}\label{author-information}

\textbf{Institute for Political Science, University of Bamberg, Bamberg,
Germany} Andreas Jungherr

\textbf{Graduate Institute of Journalism, National Taiwan University,
Taipei, Taiwan} Adrian Rauchfleisch

\newpage{}

\section{Appendix: Supplementary Materials}\label{sec-app_appendix}

\subsection{Sampling}\label{sec-app_sampling}

We queried people for (a) their interest in participating in
deliberative formats facilitated by humans or AI-enabled systems and (b)
their assessments of deliberative quality of these formats. We ran a
preregistered survey (Soft Launch n=100; Main Study n=1,850 completed
interviews) among members of an online panel that the market and public
opinion research company \emph{Ipsos} provided. Respondents from the
soft launch are not included in the analysis.

We used quotas on age, gender, region, and education to realize a sample
representative of the German electorate. As Table~\ref{tbl-quotas}
shows, the sampling was largely successful. The average interview length
was 15 minutes. The survey was fielded between February 11 and February
18, 2025. The fieldwork was conducted in compliance with the standards
ISO 9001:2015 and ISO 20252:2019, as were all study-related processes.
Before running the survey, we registered our research design, analysis
plan, and hypotheses about outcomes. We only deviate from the
preregistration in the measurement of education. Other than
preregistered, we count BA attainment or higher as high education
(Preregistration:
\url{https://osf.io/aq42e/?view_only=9f19e324effb42d9a5891ceedaf728fa}).

\begin{longtable}[]{@{}
  >{\raggedright\arraybackslash}p{(\linewidth - 6\tabcolsep) * \real{0.1087}}
  >{\raggedright\arraybackslash}p{(\linewidth - 6\tabcolsep) * \real{0.3478}}
  >{\raggedright\arraybackslash}p{(\linewidth - 6\tabcolsep) * \real{0.2935}}
  >{\raggedleft\arraybackslash}p{(\linewidth - 6\tabcolsep) * \real{0.2500}}@{}}
\caption{Comparison between official population census Germany and
realized sample (soft launch n=100 and main study
n=1,850}\label{tbl-quotas}\tabularnewline
\toprule\noalign{}
\begin{minipage}[b]{\linewidth}\raggedright
Type
\end{minipage} & \begin{minipage}[b]{\linewidth}\raggedright
Category
\end{minipage} & \begin{minipage}[b]{\linewidth}\raggedright
Official Statistics (\%)
\end{minipage} & \begin{minipage}[b]{\linewidth}\raggedleft
Realized distribution (\%)
\end{minipage} \\
\midrule\noalign{}
\endfirsthead
\toprule\noalign{}
\begin{minipage}[b]{\linewidth}\raggedright
Type
\end{minipage} & \begin{minipage}[b]{\linewidth}\raggedright
Category
\end{minipage} & \begin{minipage}[b]{\linewidth}\raggedright
Official Statistics (\%)
\end{minipage} & \begin{minipage}[b]{\linewidth}\raggedleft
Realized distribution (\%)
\end{minipage} \\
\midrule\noalign{}
\endhead
\bottomrule\noalign{}
\endlastfoot
Gender & Male & 50.0 & 50.0 \\
Gender & Female & 50.0 & 49.7 \\
Gender & Other & & 0.3 \\
Gender & NA & & 0.1 \\
Age & 18-29 Years & 18.1 & 18.1 \\
Age & 30-44 Years & 26.7 & 24.5 \\
Age & 45-59 Years & 28.7 & 28.7 \\
Age & 60-75 Years & 26.5 & 26.5 \\
Age & 76+ Years & & 2.2 \\
Region & Baden-Württemberg & 13.4 & 13.6 \\
Region & Bayern & 16.0 & 15.1 \\
Region & Berlin & 4.5 & 4.7 \\
Region & Brandenburg & 3.0 & 3.1 \\
Region & Bremen & 0.8 & 0.7 \\
Region & Hamburg & 2.3 & 2.4 \\
Region & Hessen & 7.6 & 7.2 \\
Region & Mecklenburg-Vorpommern & 1.9 & 1.8 \\
Region & Niedersachsen & 9.6 & 9.8 \\
Region & Nordrhein-Westfalen & 21.5 & 21.7 \\
Region & Rheinland-Pfalz & 5.0 & 5.1 \\
Region & Saarland & 1.2 & 1.2 \\
Region & Sachsen & 4.7 & 4.8 \\
Region & Sachsen-Anhalt & 2.6 & 2.7 \\
Region & Schleswig-Holstein & 3.5 & 3.6 \\
Region & Thüringen & 2.5 & 2.4 \\
Education & Low (ISCED 0-2) & 19.4 & 20.0 \\
Education & Medium (ISCED 3-4) & 50.6 & 49.0 \\
Education & High (ISCED 5-8) & 30.0 & 31.0 \\
\end{longtable}

\subsection{Exclusions}\label{sec-app_exclusion}

As specified in the preregistration, we used three attention checks to
identify and exclude inattentive respondents. The first was an
open-ended question, the second was hidden in an item grid, and the
third was a simple single-choice question. The three checks were
distributed throughout the entire survey. Ipsos flagged respondents if
they failed two out of three attention checks and we excluded them from
the analysis. Table~\ref{tbl-attention1} gives an overview of the number
of flagged and excluded participants.

\begin{longtable}[]{@{}ll@{}}
\caption{Number of excluded respondents, main study and soft
launch}\label{tbl-attention1}\tabularnewline
\toprule\noalign{}
Check & Number \\
\midrule\noalign{}
\endfirsthead
\toprule\noalign{}
Check & Number \\
\midrule\noalign{}
\endhead
\bottomrule\noalign{}
\endlastfoot
Check 1 (open-ended question) & 11 \\
Check 2 (item grid) & 342 \\
Check 3 (single-choice question) & 191 \\
Excluded Respondents (2 out of 3) & 102 \\
\end{longtable}

\subsection{Tasks}\label{sec-app_tasks}

We presented respondents with a random draw of a total set of 12
descriptions of deliberation tasks. These tasks were randomly attributed
to AI or humans. This set of tasks captured different elements that are
important to the successful design and execution of deliberation
formats. Table~\ref{tbl-tasks} documents types of tasks and associated
task descriptions.

\begin{longtable}[]{@{}
  >{\raggedright\arraybackslash}p{(\linewidth - 4\tabcolsep) * \real{0.1449}}
  >{\raggedright\arraybackslash}p{(\linewidth - 4\tabcolsep) * \real{0.4638}}
  >{\raggedright\arraybackslash}p{(\linewidth - 4\tabcolsep) * \real{0.3913}}@{}}
\caption{Deliberative tasks.}\label{tbl-tasks}\tabularnewline
\toprule\noalign{}
\begin{minipage}[b]{\linewidth}\raggedright
Type
\end{minipage} & \begin{minipage}[b]{\linewidth}\raggedright
Task
\end{minipage} & \begin{minipage}[b]{\linewidth}\raggedright
Description
\end{minipage} \\
\midrule\noalign{}
\endfirsthead
\toprule\noalign{}
\begin{minipage}[b]{\linewidth}\raggedright
Type
\end{minipage} & \begin{minipage}[b]{\linewidth}\raggedright
Task
\end{minipage} & \begin{minipage}[b]{\linewidth}\raggedright
Description
\end{minipage} \\
\midrule\noalign{}
\endhead
\bottomrule\noalign{}
\endlastfoot
Recruitment \& Selection & Balanced Participation & {[}AI tools/Human
moderators{]} select participants for online discussions, ensuring
demographic diversity and balancing power dynamics to make sure many
different views are heard and given equal weight. \\
& Topic Identification & {[}AI tools/Human moderators{]} identify topics
for online discussions by looking for trending issues, recurring
concerns, and emerging themes, ensuring the topics are relevant and
representative of public interests. \\
Information and argument processing & Content Summarization & {[}AI
tools/Human moderators{]} can summarize contributions and information to
help participants gain an overview and assess the available content. \\
& Argument Highlighting & {[}AI tools/Human moderators{]} can highlight
key arguments to promote critical thinking and emphasize essential
discussion points. \\
& Argument Structuring & {[}AI tools/Human moderators{]} can map
arguments and detect areas of agreement to guide discussions toward
collaborative decision-making. \\
& Opinion Aggregation & {[}AI tools/Human moderators{]} can aggregate
opinions and preferences from participants to reflect collective
decisions effectively and transparently. \\
Discussion moderation & Fact Checking & {[}AI tools/Human moderators{]}
can fact-check claims and summarize material to provide participants
with accurate and accessible information. \\
& Disruption Detection & {[}AI tools/Human moderators{]} can maintain
respect and balance in conversations by flagging disruptive behavior and
promoting inclusivity. \\
& Tone \& Authenticity & {[}AI tools/Human moderators{]} can analyze
sentiment to assess discourse authenticity and use discourse mapping to
ensure a plurality of viewpoints and inclusive participation. \\
Discussion facilitation & Gamification & {[}AI tools/Human moderators{]}
can emphasize selected arguments or use gamification features to
encourage critical thinking and help participants engage more deeply
with evidence and reasoning. \\
& Translation Support & {[}AI-powered translation tools/Human
translators{]} can help participants engage in discussions across
language barriers, making deliberation more inclusive. \\
& Role-play Empathy & {[}AI-driven role-playing agents can
simulate/Human moderators can provide{]} alternative perspectives,
helping participants critically engage with opposing arguments. \\
\end{longtable}

\subsection{Measures}\label{sec-app_measures}

We show the English translations of the questions in the following
table. The complete questionnaire with all questions in German is
available in our preregistration on OSF:
\url{https://osf.io/aq42e/?view_only=9f19e324effb42d9a5891ceedaf728fa}.
In the following table, all items marked with (-) are negatively
formulated and recoded for the positive index. The mean scores for the
index are always based on the recoded items.

\begin{landscape}
\begin{longtable}{p{2cm}p{3cm}p{7cm}p{5cm}cc}
\caption{Descriptive statistics for all relevant variables and items. The Cronbach's $\alpha$ for Deliberative quality were calculated individually for each case and moderator combination.}
\label{tab:descr}\\
\toprule
& Variable & Question & Operationalization & M (SD) & n\\
\midrule
\endfirsthead

\toprule
& Variable & Question & Operationalization & M (SD) & n\\
\midrule
\endhead

\multicolumn{6}{r}{{ }} \\
\endfoot
    
\bottomrule
\endlastfoot

\toprule

RQ1/H1-H7& Willingness to participate in deliberation & If you had the chance to participate in such a session, how interested do you think you would be in doing so? & \begin{itemize}\item 1="Not at all interested"\item 7="Extremely interested\end{itemize} & 3.95 (1.98) & 11100\\

RQ2/H8-H12& Deliberative quality (5 items, $\alpha$ = 0.73-0.84) & Please indicate how much you agree with the following statements when you think about the case we have just described & \begin{itemize}\item 1="Completely Disagree"\item 7="Completely Agree"\end{itemize} & 4.16 (1.29) & 11100\\
 && This makes it difficult for people like me to get their voices heard. (-) && 4.31 (1.75) & 11100\\
 && This makes it easy to identify the best policy solution. && 4.02 (1.67) & 11100\\
 && This ensures a fair discussion and policy process. && 4.19 (1.67) & 11100\\
 && I trust this process. && 4.01 (1.76) & 11100\\
 && This will give space for a plurality of diverse views on the issue. && 4.28 (1.64) & 11100\\

H1a& Income & What is the total monthly net income (after taxes) of your household, earned by all household members? & \begin{itemize}\item 1=€0-€500\item 2=€501-€750\item 3=€751-€1000\item 4=€1001-€1250\item 5=€1251-€1500\item 6=€1501-€1750\item 7=€1751-€2000\item 8=€2001-€2500\item 9=€2501-€3000\item 10=€3001-€4000\item 11=€4001-€5000\item 12=€5001-€10.000\item 13=€10.001 and more\end{itemize} & 8.52 (2.85) & 1716\\
H1a& Employment, full time & & 1 = Employed full-time (30 hours or more per week) & 0.49 & 1823\\
H1a& Children, 12 or younger & & Count & 0.27 (0.82) & 1721\\
H1a& Education & & 1 = University degree incl. Bachelor & 0.23 & 1850\\
H1a& Past participatory activity & Over the past two years, have you been a member or participated in activities of...\begin{itemize}\item"service club or fraternal organizations (e.g., Elks, Rotary)"\item"veterans groups"\item "religious groups"\item "senior citizen's centers or groups"\item "women's groups"\item "issue-oriented political organizations"\item "political parties"\item "non-partisan civic organizations"\item "school clubs or associations"\item "hobby, sports teams, or youth groups"\item "neighborhood associations or community groups"\item "groups representing racial/ethnic interests"\item "Other civil society groups (please specify)"\end{itemize} & Sum index: Yes/No  & 0.68 (1.08) & 1641\\
H1a& Past deliberative activity & Over the past two years, have you participated in... \begin{itemize}\item"Town Hall Meetings: Open forums where community members discuss specific issues with policymakers, often including a Q\&A session."\item "Deliberative Polling: Participants discuss an issue in small groups after being provided with relevant information, with opinions polled before and after deliberation."\item "World Café: Small, rotating discussion groups explore specific topics, with insights shared in a plenary session."\item "Citizens' Assemblies: Randomly selected citizens deliberate on an issue over multiple sessions, often resulting in formal recommendations."\item "Online Town Halls: Virtual forums where participants engage with policymakers or leaders via live video, chat, or Q\&A tools."\item "Virtual Roundtables: Moderated discussions held via video conferencing platforms, such as Zoom or Microsoft Teams."\item "Massive Open Online Deliberation (MOOD): Large-scale platforms enabling asynchronous, in-depth deliberation on complex issues."\item "Structured Online Discussions: Organized and moderated conversations on platforms such as Twitter, Reddit, or Facebook, designed to encourage thoughtful and focused dialogue, often guided by hashtags or discussion facilitators."\item "Other deliberative formats (please specify)"\end{itemize} & Sum index: Yes/No  & 0.39 (0.90) & 1715\\

H1b& Political interest & I am interested in politics. & \begin{itemize}\item 1="Completely Disagree"\item 7="Completely Agree"\end{itemize} & 4.91 (1.77) & 1850\\
H1b& Political efficacy (2 items, Spearman-Brown = 0.67) & & \begin{itemize}\item 1="Completely Disagree"\item 7="Completely Agree"\end{itemize} & 3.34 (1.57) & 1850\\
 & &Public officials don’t care what people like me think. (-) && 4.94 (1.77) & 1850\\
 & &I can’t influence government decisions. (-) && 4.38 (1.85) & 1850\\
H1b& PID Strength & First selecting PID: "In Germany, many people tend to align with a particular political party for a longer period, even though they occasionally vote for a different party. How about you: Do you, generally speaking, tend to align with a specific party? And if so, which one?" & \begin{itemize}\item"(1) SPD"\item "(2) CDU"\item "(3) CSU"\item "(4) GRÜNE"\item "(5) FDP"\item "(6) AfD"\item "(7) DIE LINKE"\item "(8) Bündnis Sahra Wagenknecht (BSW)"\item "(9) andere Partei, und zwar"\item "(10) keiner Partei"\end{itemize} &  & \\
&  & Then "How strongly or weakly do you, all things considered, align with this party?" & \begin{itemize}\item 1 Very weakly - \item 7 Very strongly\end{itemize}. Respondents with the response "keiner Partei/no party" were coded as "1 Very weakly" & 4.55 (1.95) & 1850\\

H2& Need for cognition (4 items, $\alpha$ = 0.9) & & \begin{itemize}\item 1="Completely Disagree"\item 7="Completely Agree"\end{itemize} & 4.72 (1.38) & 1850\\
 && I find satisfaction in deliberating hard and for long hours. && 4.68 (1.58) & 1850\\
 && The notion of thinking abstractly is appealing to me. && 4.65 (1.64) & 1850\\
 && I really enjoy a task that involves coming up with new solutions to problems. && 4.72 (1.57) & 1850\\
 && I like tasks that require much thought and mental effort. && 4.83 (1.50) & 1850\\

H3a/H8a& Technology benefits (3 items, $\alpha$ = 0.86) && \begin{itemize}\item 1="Completely Disagree"\item 7="Completely Agree"\end{itemize} & 4.63 (1.38) & 1850\\
 && I feel more in control of my daily activities thanks to technology. && 4.46 (1.61) & 1850\\
 && Technology helps me manage my time more effectively. && 4.77 (1.52) & 1850\\
 && I feel more productive in both my personal and professional life because of technology. && 4.65 (1.58) & 1850\\

H3b/H8b& Technology risks (3 items, $\alpha$ = 0.76) && \begin{itemize}\item 1="Completely Disagree"\item 7="Completely Agree"\end{itemize} & 3.65 (1.43) & 1850\\
 && I feel more in control of my daily activities thanks to technology. && 3.26 (1.70) & 1850\\
 && Technology helps me manage my time more effectively. && 4.10 (1.77) & 1850\\
 && I feel more productive in both my personal and professional life because of technology. && 3.58 (1.75) & 1850\\

H4/H9& AI benefits (5 items, $\alpha$ = 0.87) && \begin{itemize}\item 1="Completely Disagree"\item 7="Completely Agree"\end{itemize} & 3.76 (1.35) & 1850\\
 && AI will drive significant economic expansion in this country. && 3.87 (1.60) & 1850\\
 && AI will provide our military with advanced defense capabilities, ensuring national security. && 3.88 (1.72) & 1850\\
 && AI will help governments to more efficiently plan for the future and manage crises. && 3.67 (1.67) & 1850\\
 && AI helps parties to communicate more successfully with voters. && 3.62 (1.72) & 1850\\
 && AI will help humanity to address existential threats more successfully. && 3.76 (1.63) & 1850\\

H5/H10& AI risks (5 items, $\alpha$ = 0.83) & &\begin{itemize}\item 1="Completely Disagree"\item 7="Completely Agree"\end{itemize} & 4.58 (1.33) & 1850\\
 && AI is likely to cause widespread job displacement and unemployment. && 4.16 (1.73) & 1850\\
 && Unchecked AI development could pose existential threats to humanity. && 4.91 (1.72) & 1850\\
 && AI in military applications can lead to unintended escalations or conflicts due to lack of human judgment. && 4.47 (1.67) & 1850\\
 && As AI increasingly takes over decision-making, we risk losing control over our lives. && 4.78 (1.75) & 1850\\
 && AI will allow parties to manipulate voters more successfully. && 4.61 (1.76) & 1850\\

H6/H11& AI experience (2 items, $\alpha$ = 0.77) & How frequently do you use AI supported applications or services in your... &\begin{itemize}\item 1="never"\item 7="very often"\end{itemize} & 2.72 (1.66) & 1850\\
 && professional or work environment && 2.47 (1.84) & 1850\\
 && personal life and spare time && 2.96 (1.83) & 1850\\

H7/H12& AI anthropomorphism (4 items, $\alpha$ = 0.88) & &\begin{itemize}\item 1="Completely Disagree"\item 7="Completely Agree"\end{itemize} & 3.18 (1.54) & 1850\\
 && AI has the potential to develop a sense of humor. && 3.32 (1.80) & 1850\\
 && An AI can have a unique personality. && 3.19 (1.81) & 1850\\
 && AI has the capacity to develop consciousness. && 3.24 (1.80) & 1850\\
 && AI can develop a sense of morals and ethics. && 2.98 (1.75) & 1850\\

\end{longtable}
\end{landscape}

\subsection{Validation}\label{sec-cfa}

In a pre-test with a small sample (n = 105), we examined the performance
of the items for the scales \emph{Technology Risk} and \emph{Technology
Benefits} using an exploratory factor analysis (EFA). The item wordings
in the tables are provided in English. For the original German wording
of the items used in the survey, please refer to the questionnaire in
the preregistration
(\url{https://osf.io/aq42e/?view_only=9f19e324effb42d9a5891ceedaf728fa}).

\begin{table}[H]
    \centering
    \begin{tabular}{p{0.55\textwidth} c c}
    \hline
    \textbf{Item Wording} & \textbf{Benefits} & \textbf{Risks} \\
    \hline
    I feel more in control of my daily activities thanks to technology. & 0.83 & \\[0.5em]
    Technology helps me manage my time more effectively. & 0.78 & \\[0.5em]
    I feel more productive in both my personal and professional life because of technology. & 0.83 & \\[0.5em]
    Technology increases my stress levels rather than making life easier. &  & 0.78 \\[0.5em]
    Technology reduces the quality of face-to-face interactions with others. &  & 0.58 \\[0.5em]
    Technology contributes to distractions and reduces my ability to focus. &  & 0.79 \\
    \hline
    \end{tabular}
    \caption{Factor Loadings from the exploratory factor analysis with oblimin rotation. Factor loadings smaller than $\pm$0.1 are not shown.}
    \label{tab:efa_loadings}
\end{table}

Here, we present the results of a confirmatory factor analysis. In
general, the CFA indicates a good fit for the model with values above or
below the generally recommended fit indicators in the literature (CFI =
0.987, TLI = 0.975, RMSEA = 0.061, SRMR = 0.03). Although the chi-square
test is significant (\(\chi^2\)(8, N = 1850) = 63.63, p \textless{}
.001), this is expected given the large sample size (n = 1850), as the
test is highly sensitive to sample size. The following table presents
the standardized factor loadings from the CFA.

\begin{table}[H]
    \centering
    \begin{tabular}{p{0.55\textwidth} c c}
    \hline
    \textbf{Item Wording} & \textbf{Benefits ($\lambda$)} & \textbf{Risks ($\lambda$)} \\
    \hline
    I feel more in control of my daily activities thanks to technology. & 0.809 & \\[0.5em]
    Technology helps me manage my time more effectively. & 0.801 & \\[0.5em]
    I feel more productive in both my personal and professional life because of technology. & 0.837 & \\[0.5em]
    Technology increases my stress levels rather than making life easier. &  & 0.812 \\[0.5em]
    Technology reduces the quality of face-to-face interactions with others. &  & 0.607 \\[0.5em]
    Technology contributes to distractions and reduces my ability to focus. &  & 0.740 \\
    \hline
    \end{tabular}
    \caption{Standardized Factor Loadings for Technology’s Benefits and Risks in Daily Life from the confirmatory factor analysis. The correlation between the two factors is -0.33.} 
    \label{tab:tech_loadings}
\end{table}

\subsection{Model results}\label{sec-app_models}

As specified in the preregistration under \emph{Missing data}, we used
data imputation to fill in missing responses for:

\begin{itemize}
\tightlist
\item
  past participatory activity (sum index),
\item
  past deliberative activity (sum index),
\item
  full-time employment (1-fully employed vs 0-other),
\item
  number of children in the household aged 12 or younger, and
\item
  income.
\end{itemize}

For data imputation, we followed the procedure recommended in the
literature (van Buuren, 2012/2018). Using the R package \emph{mice} (van
Buuren \& Groothuis-Oudshoorn, 2011), we created 100 datasets with
imputed data for the missing values using predictive mean matching (van
Buuren, 2012/2018). We used the values of the variables mentioned above
(if available), all the measured independent variables, age, and gender
(male), as predictors for predictive mean matching. After imputing the
data, we estimate the model (multilevel model with varying intercept for
cases and participants) for each dataset and pooled the results also
with the mice package in R (Barnard \& Rubin, 1999; Rubin, 1987).

\begin{table}[H]
\centering
\begin{tabular}{lccc}
\toprule
\textbf{Predictors} & \textbf{Estimates} & \textbf{CI} & \textbf{\textit{p}} \\ 
\midrule
Intercept & 2.622 & [2.219, 3.026] & 0.000 \\
RQ1a: Human vs AI moderation & -0.296 & [-0.336, -0.255] & 0.000 \\

H1a: Income & 0.005 & [-0.020, 0.031] & 0.692 \\
H1a: Employment, full time & 0.156 & [0.016, 0.297] & 0.030 \\
H1a: Children, 12 or younger & -0.007 & [-0.089, 0.075] & 0.868 \\
H1a: Education & 0.080 & [-0.075, 0.234] & 0.312 \\
H1a: Past participatory activity & 0.007 & [-0.071, 0.086] & 0.853 \\
H1a: Past deliberative activity & 0.050 & [-0.045, 0.146] & 0.304 \\

H1b: Political interest & 0.125 & [0.084, 0.167] & 0.000 \\
H1b: Political efficacy & 0.025 & [-0.016, 0.067] & 0.234 \\
H1b: PID strength & 0.017 & [-0.018, 0.051] & 0.340 \\

H2: Need for cognition & 0.199 & [0.149, 0.250] & 0.000 \\

H3a: Technology benefits X Treatment & 0.002 & [-0.034, 0.038] & 0.915 \\
H3b: Technology risks X Treatment & -0.003 & [-0.035, 0.030] & 0.861 \\

H4: AI benefits X Treatment & 0.074 & [0.034, 0.113] & 0.000 \\
H5: AI risks X Treatment & -0.125 & [-0.158, -0.093] & 0.000 \\

H6: AI experience X Treatment & 0.035 & [0.005, 0.065] & 0.023 \\
H7: AI anthropomorphism X Treatment & 0.100 & [0.066, 0.134] & 0.000 \\

Technology benefits & 0.213 & [0.154, 0.272] & 0.000 \\
Technology risks & 0.020 & [-0.033, 0.073] & 0.462 \\
AI benefits & 0.200 & [0.135, 0.264] & 0.000 \\
AI risks & 0.075 & [0.022, 0.128] & 0.005 \\
AI experience & 0.115 & [0.062, 0.167] & 0.000 \\
AI anthropomorphism & 0.002 & [-0.052, 0.056] & 0.938 \\
Gender (male = 1) & -0.069 & [-0.199, 0.062] & 0.303 \\
Age & -0.008 & [-0.013, -0.003] & 0.001 \\
\bottomrule
\end{tabular}
\caption{Result based on the pooled models with varying intercepts for cases (12) and participants (1850) with a total of 11000 observations. The dependent variable in this model is Willingness to participate in deliberation.}
\end{table}

\begin{table}[H]
\centering
\begin{tabular}{lccc}
\toprule
\textbf{Predictors} & \textbf{Estimates} & \textbf{CI} & \textbf{\textit{p}} \\ 
\midrule
Intercept & 3.775 & [3.541, 4.010] & 0.000 \\
RQ1b: Human vs AI moderation & -0.293 & [-0.324, -0.263] & 0.000 \\

H8a: Technology benefits X Treatment & -0.011 & [-0.038, 0.016] & 0.430 \\
H8b: Technology risks X Treatment & 0.002 & [-0.022, 0.027] & 0.843 \\
H9: AI benefits X Treatment & 0.104 & [0.074, 0.133] & 0.000 \\
H10: AI risks X Treatment & -0.139 & [-0.164, -0.115] & 0.000 \\
H11: AI experience X Treatment & 0.004 & [-0.019, 0.026] & 0.759 \\
H12: AI anthropomorphism X Treatment & 0.118 & [0.092, 0.143] & 0.000 \\

Income & 0.008 & [-0.007, 0.022] & 0.304 \\
Employment, full time & 0.069 & [-0.010, 0.148] & 0.087 \\
Children, 12 or younger & -0.011 & [-0.057, 0.034] & 0.628 \\
Education & 0.029 & [-0.058, 0.116] & 0.512 \\
Past participatory activity & -0.007 & [-0.050, 0.036] & 0.755 \\
Past deliberative activity & -0.014 & [-0.068, 0.040] & 0.607 \\

Political interest & 0.035 & [0.012, 0.059] & 0.003 \\
Political efficacy & 0.041 & [0.017, 0.065] & 0.001 \\
PID strength & 0.009 & [-0.010, 0.028] & 0.363 \\

Need for cognition & 0.080 & [0.052, 0.108] & 0.000 \\

Technology benefits & 0.162 & [0.127, 0.196] & 0.000 \\
Technology risks & -0.011 & [-0.042, 0.020] & 0.493 \\
AI benefits & 0.249 & [0.211, 0.287] & 0.000 \\
AI risks & 0.024 & [-0.007, 0.055] & 0.129 \\
AI experience & 0.004 & [-0.027, 0.034] & 0.817 \\
AI anthropomorphism & -0.032 & [-0.064, -0.001] & 0.045 \\
Gender (male = 1) & -0.115 & [-0.188, -0.041] & 0.002 \\
Age & -0.005 & [-0.008, -0.002] & 0.000 \\
\end{tabular}
\caption{Result based on the pooled models with varying intercepts for cases (12) and participants (1850) with a total of 11000 observations. The dependent variable in this model is Perceived Deliberative Quality.}
\end{table}

\newpage{}

\section*{References}\label{references}
\addcontentsline{toc}{section}{References}

\phantomsection\label{refs}

\end{document}